# Virtual Proximity Citation (VCP): A Supervised Deep Learning Method to Relate Uncited Papers On Grounds of Citation Proximity

## Rohit Rawat

### School of Computer Science & Statistics, Trinity College Dublin, Ireland

## ABSTRACT

Citation based approaches have seen a good progress for recommending research papers using citations in the paper. Citation proximity analysis which uses the in-text citation proximity to find relatedness between two research papers is better than co-citation analysis and bibliographic analysis. However, one common problem which exist in each approach is that paper should be well cited. If documents are not cited properly or not cited at all, then using these approaches will not be helpful. To overcome the problem, this paper discusses about the approach Virtual Citation Proximity (VCP) which uses Siamese Neural Network along with the notion of citation proximity analysis and content-based filtering. To train this model, actual distance between the two citations in a document is used as a ground truth, this distance is basically the word count between the two citations. VCP is trained on a Wikipedia articles for which the actual word count is available which is used to calculate the similarity between the documents. This can be used to calculate relatedness between two documents in a way they would have been cited in the proximity even if the documents are uncited. This approach has shown a great improvement in predicting proximity with basic neural networks over the approach which uses Average Citation Proximity index value as the ground truth. This can be improved by using complex neural network and proper hyper tuning of parameters.

## 1 Introduction

Research paper recommender system has seen emerging growth in terms of techniques used to recommend similar paper. Many research papers have been published in this domain [1] to discuss various recommendation approach. However, citation-based approach has proven to be more significant in recommending research papers. Citations based approaches can be Bibliographic coupling and Co-citation analysis. Bibliographic coupling approach uses a coupling strength which means if two document cites one or more common document, then they are similar [2]. Problem with Bibliographic coupling method is it refer to the bibliography section of the documents and many times it contains references which is not being used in the document, it is called false citations [3]. On the other hand, co-citation analysis approach uses citation to find similarity in documents. If two documents have been cited in one or more paper, then they are similar [4]. Problem with co-citation analysis is that, it neither considers the content of the document nor the position of the citations. This can lead to inconsistency in the result; if two papers are cited in two different section of the document for different reasons will be considered similar. To overcome above problems, Citation Proximity Analysis [5] is one of the promising approaches. This approach uses in-text citation to calculate the citation proximity index (CPI). That means if citations occur in same line, it will have higher CPI value and thus the papers will be similar compared to citations occurring in different paragraphs or sections, the same is illustrated in the Figure 1. CPA is considered to perform better than other citations approaches as per the experiment conducted by Schwarzer M [7] on the Wikipedia articles.

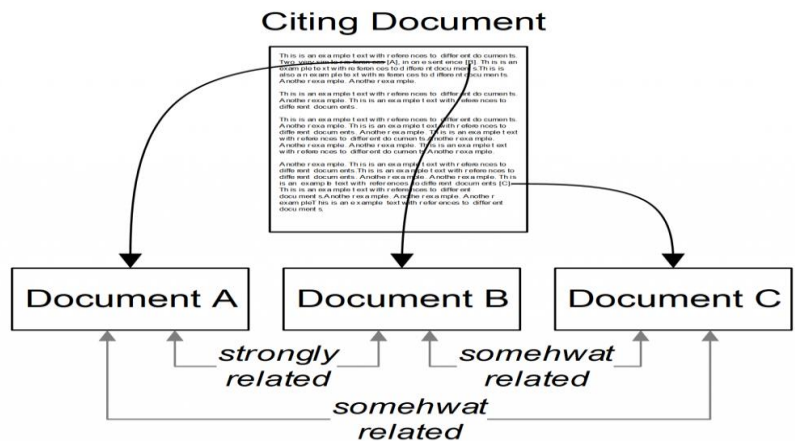

**Figure 1: Citation Proximity Analysis [5]. It illustrates document A and B are strongly related as they are cited in same line. On the other hand, document A and C, as well as document B and C are somewhat related as they are cited in different sections**.

Apart from recommending research papers and its use in Wikipedia articles, CPA has been used in webpages as link citation proximity [8]. Like other citation approaches, CPA has its own disadvantages, it also does not consider the content of the cited documents; two documents are cited because they are addressing same problem but using different method will be considered similar. Another limitation of this approach is that, it cannot be used for recommending the research paper which is not cited. In general, for a research paper to get citation it takes more than a year [11,12]. Vincent L [13] also concluded from his experiments that researchers mostly do not rely on the recent research which is one of the reasons some paper go uncited even after many years of publication. To address this problem, a novel approach has been proposed by Beel J [6], Virtual Citation Proximity (VCP). VCP, in general, uses power of deep learning on top of citation proximity analysis along with the advantage of content-based filtering to predict the relatedness between the two documents, even if they are uncited. VCP can improve the current recommendation approach to suggest research papers significantly using machine learning.

## 1.1 Virtual Citation Proximity

Virtual citation proximity is used to predict the relatedness between two documents in the proximity that they would have been cited if they are ever cited. VCP takes into consideration the content of the two documents and compare it based on the real ground truth value which is the distance between the two co-cited documents. This distance is calculated as the count of the number of words between them. This way VCP will be able to predict the proximity like the citation proximity analysis. VCP makes use of artificial neural network to compare two documents based on CPA analysis and content-based filtering. Siamese neural network is used to compare two documents. For each document, a similar neural network is designed which is then inserted into Siamese network along with ground truth value which is the number of words occurring between the citations in our implementation. This neural network tries to learn the relatedness like the CPA and predicts the similarity on the same basis. The inputs to this neural network will be two documents which is basically the text (abstract). These texts are converted into word embeddings, which is nothing but a vector which will make learning easy for the neural networks as it understands number. Once the neural network is trained it will be able to predict the relatedness between two documents which will be like the CPA and if it is accurate enough then it will change the way of recommendation system effectively. Unlike CPA, VCP considers all the documents whether they are cited or not and thus its range is much better than CPA. It is not necessary for both the documents to have common terms as neural network is studying the underlying features based on CPA analysis as well. Molloy P [14] VCP implementation achieved a MAE of 0.0055 which is 20% better than mean baseline evaluating neural network on 2 million co-cited articles from Wikipedia. Dataset developed by Molloy P [14] included Citation Proximity index (CPI) which is calculated based on the position of the citations. It is developed by using tool Citolytics [7]. Molloy P [14] first rules out the co-cited articles which are cited less than 5 times. Second, Average CPI is considered as a ground truth for the learning purpose. Another thing, Molloy P [14] has used 200 maximum sequence length for the article. If co-cited articles with multiple citations are considered with average CPI as ground truth, this will not help neural network to learn how two documents should be cited in proximity in an individual paper. Apart from that it introduces bias in the data, like if articles are cited in same line in one article then it will improve the overall CPI value. This cannot be used to make predictions correctly. Molloy P [14] considered more the number of words in text better will be the prediction, but to achieve this maximum number of articles should have stated maximum number of words else the embeddings will consist mostly vectors of 0 which will not help neural network to learn anything instead it will make neural network heavy and that is why after first epoch, learning curve did not show any improvement. To improve on this, main goal of this research would be to focus on co-cited articles which are cited only once, to remove the bias ingredient from it. And secondly, to consider actual distance i.e. count of the words between two citations as a ground truth for the neural network to train. Dataset for this is the same as used by Molloy P [14]. The result obtained using our approach has shown a significant improvement over the Molloy P [14] approach although the neural network is trained on 64000 co-cited articles only. MAE of 263.16 is obtained on 20000 test data which is 38% better than the average baseline.

## 2  Related Work

Virtual Citation Proximity is a novel approach in calculating relatedness between uncited documents using proximity distance as a ground truth using artificial neural network and there is no other approach like this or solving this problem the same. However, expert's judgement is being used as a ground truth in MeSH, ACM, ACL. MeSH classification created by medical experts to classify major field and sub field in a classification tree. MeSH indexing is also like VCP except it is classified by human indexers [9]. It is mostly used in biomedical research and it costly as well as time consuming.  Once the classification is done, documents belonging to same label are related and later this can be used in recommendation system. MeSH Now [10] is the state of the art used to index the article automatically. Classification is done at the time of publication, so if classification scheme changes this will not be updated. Virtual citation proximity is free from these problems as it is not subjected to single domain and there is no need to classify any paper. It simply uses deep learning to calculate relatedness between paper which can be extended to every domain.

## 3  Methodology

### 3.1  Overall Design

VCP approach considers comparing two documents which is a text and using count of the number of words between citations as ground truth. To achieve this, Siamese Neural Network [15]is the best approach as it precisely compare two objects/documents. Siamese neural network consists of a dual branch with shared weights. Both the branches are exact replica of each other. The output from these branches are combined to calculate the similarity between two branches. Similarly, this can be used in comparing documents where two branches are the co-cited articles which is combined to find relatedness which is calculated based on the word count between two citations LSTM (Long Short Term Memory) is used in each branch to learn from the text. LSTM is very successful in text classification [16] and it is used to generate semantic meaning in the articles which will be very useful while comparing two articles. Before feeding text into LSTM layer, embedding layer is used to create a word vectors for the input using glove 840b 300d [17]. Figure 2 shows the overall design of the implementation. The co-cited articles are embedded into a word vector which then acts as an input to LSTM layer. LSTM layer learns underlying features in text which is like CPA and then the output from two LTSM layers are combined to learn the comparison, after which the network is passed through the dense layer which is an output layer. Throughout the process, the weights are shared between the left and right branches. Source code is available at : https://github.com/rawatr2003/Virtual-Citation-Proximity

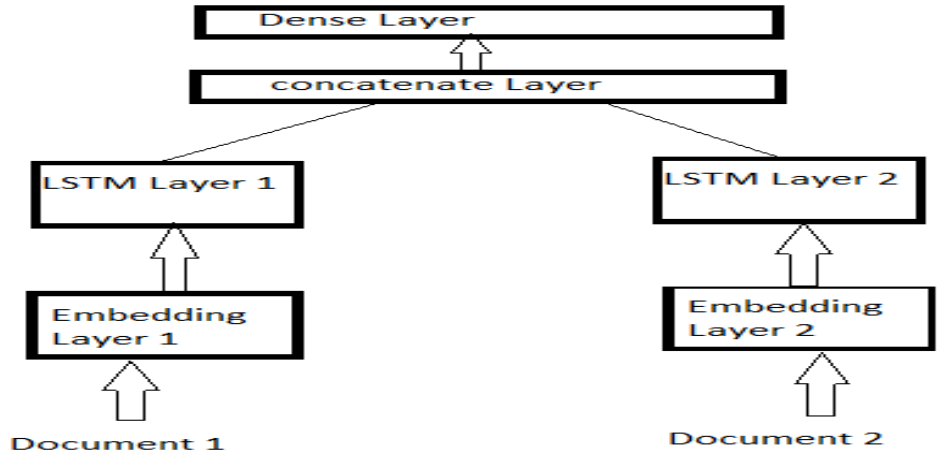

*Figure 2: Siamese Neural Network Architecture. The weights are shared between the two branches.*

## 3.2 Dataset

To train VCP neural network, data of co-cited articles along with the citation proximity between them are needed. Wikipedia consist of millions of articles which can be useful for the creation of dataset as its hyperlink can be considered as citations. Wikipedia 1st Jan 2019 dumps have been used to create a dataset which consists of title of the articles, its texts, the distance between two citations, count and the CPI value. Citolytics tool [7] is used to calculate the distance, count and the CPI value. This dataset was created by Molloy P [14] as a part of the research. However, for this research purpose, only the articles which are co-cited only once will be considered and the ground truth value is going to be the distance between the two citations i.e. the count of the words between them. This will be removing the inconsistency in the dataset and help neural network to learn how two uncited articles can be cited in the proximity. Apart from that, we are considering the co-cited articles for which the distance between them is less or equal to 1000 words, as we want to learn the similarity for which the distance is less. Figure 3 shows the distribution of distance between the co-cited articles.

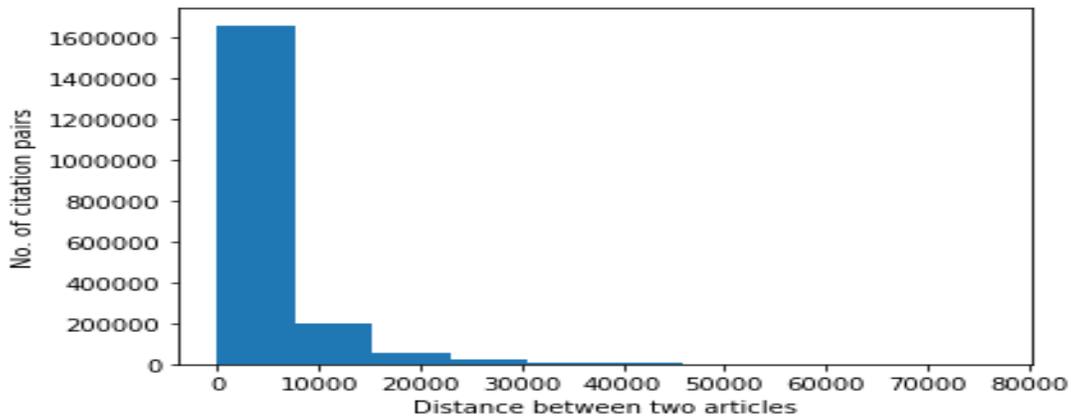

*Figure 3: Distribution of the distance between citations in the dataset.*

### 3.3 Implementation

To implement this, we give two co-cited articles as input which has been pre-processed by removing all the English stop words and all the documents have been lemmatize. Then these documents are tokenized to convert each word with the number by the frequency they have occurred in the corpus. Each tokenized document is then represented only by 50 words, as most of the documents have around 50 words, this will help neural network to learn the underlying features. Figure 4 shows the distribution of articles as per the number of words it has. Most of the articles have total word count around 50, so having 200 features will just make model worse.

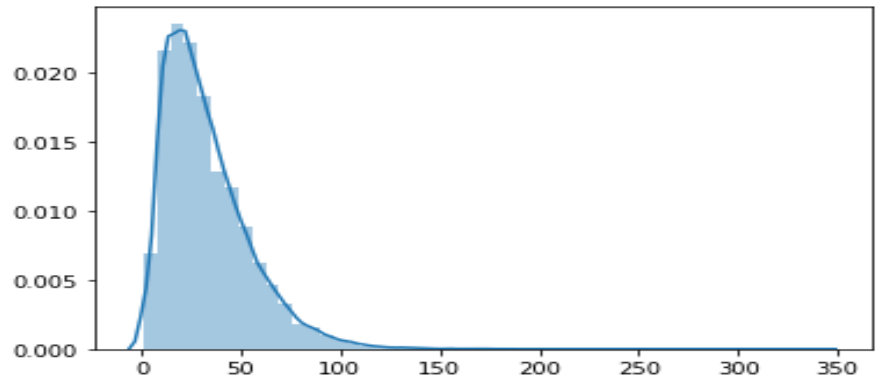

*Figure 4 Distribution of articles having number of words.*

These documents with 50 features are embedded using embedding layer. Glove 840b 300d is used to create an embedding matrix, based on which embedding layer create word vectors of 300 dimensions. These embedded documents are passed into LSTM network with 50 neurons followed by a flatten layer. The two networks are combined and fed into a dense layer which is the output layer. To avoid overfitting, early stopping is used, and this network is trained on 100 epochs. For the purpose of baseline, it can be compared with the performance of the Molloy P [14] VCP approach, comparing to which our approach has shown better improvement. Apart from that, it can be compared with the average baseline i.e. the average of the distance between the co-cited articles in the corpus.

### 3.4 Evaluation Metrics

Mean absolute error is used as an evaluation metrics over RMSE which is sensitivity to the outliers and its value changes with the sample size. Range of the distance goes from 1 to 1000 for the purpose of this experiment, so the prediction of the VCP approach will be between this depending on the similarity of the article. Documents are similar if the distance is less and are more likely to be cited together in the near proximity if they would have been cited in real. Other recommender system-based evaluation metrics can also be used to make this experiment robust which is the future scope of this research.

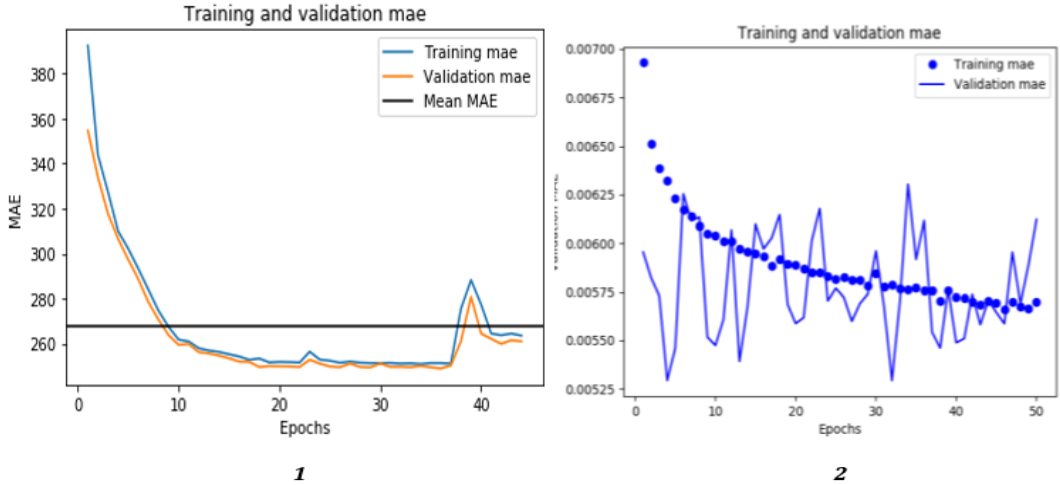

*Figure 5 MAE of the network over the epochs. 1 is our implementation. 2 is Molloy P's [14] implementation.*

## 3.5 Result and Conclusion

Our simple Siamese LSTM neural network was able to learn the underlying features based on the citation proximity to predict the relatedness of the uncited articles in the near proximity. As our network was supposed to be trained on 100 epochs, but after 44[th] epoch it stopped to avoid overfitting as early stopping is being used. Figure 5 shows the mean absolute error of the network for our implementation as well as Molloy P's [14] implementation for the comparison purpose. It is clear from figure 5 that our neural network learnt better than the Molloy P's [14] implementation, as the validation MAE curve for Molloy's implementation is very distorted. The scale of MAE value is different as both implementations have used different features for ground truth value. On evaluating the test data, MAE of 263.16 was achieved which is better than the average baseline and the average MAE of the training. Our approach performed very well on the test data with MAE of 263.16 which is 38% better than the average baseline to that 20% of Molloy P [14] when compared to the average baseline. The result of our implementation is better than Molloy's approach when compared to base average line and it removes the biasness of the data by considering documents which are only co-cited once. However, this can be improved as this training was done only on 64000 co-cited articles. By using more data and more complex neural network, the error can be minimized further. Dropout can be used to further remove the overfitting of the data. Attention mechanism can be used as a technique to give more importance to some words as per its use/importance in the sentence. Like most of the research paper has keyword section, if that can be involved in the text, it will be useful as it may contain the similar keywords which can be used to give more weightage to the co-cited articles. Different neural network can be used to train the VCP, to see if it decreases the MAE. Currently, we considered distance less than equal to 1000, we can try with less than 500 or 100 proximity distance to check the functioning of the neural network. Trained neural network can predict distance in proximity like the CPA which is a good start. However, the result is not compared with the other citation-based method used for recommending research paper which is a next task. We have not considered any other recommender system related evaluation metric, which is the important step in making VCP approach a proper VCP based recommender system. VCP can outperform citation-based approaches for calculating relatedness between the articles as it considers both the content and the citation proximity which is unique.

# 4 ACKNOWLEDGMENTS

I would like to thank my Supervisor Dr. Joeran Beel for guiding and helping me throughout the research project. I would like to thank Paul Molloy for providing me with the general dataset.